\documentclass[english,groupedaddress, superscriptaddress,pre,10pt,floatfix,reprint,twocolumn]{revtex4-1}
\usepackage[T1]{fontenc}
\usepackage[utf8]{inputenc}
\usepackage[normalem]{ulem}
\usepackage{amsmath,amsthm,mathtools,amssymb}
\usepackage{graphicx}

\usepackage{longtable}
\usepackage{tabularx}
\usepackage{appendix}
\usepackage{listings}

\usepackage{units}

\usepackage{url}

\usepackage[breaklinks]{hyperref}

\usepackage[usenames,dvipsnames,svgnames,table]{xcolor}

\begin{document}

\title{Revealing interactions between HVDC cross-area flows and frequency stability with explainable AI}

\author{Sebastian P\"utz}
\email{s.puetz@fz-juelich.de}
\affiliation{Forschungszentrum J\"ulich, Institute for Energy and Climate Research - Systems Analysis and Technology Evaluation (IEK-STE), 52428 J\"ulich, Germany}
\affiliation{Institute for Theoretical Physics, University of Cologne, 50937 K\"oln, Germany}
\author{Benjamin Sch\"afer}
\affiliation{Karlsruhe Institute of Technology, Institute for Automation and Applied Informatics (IAI),76344 Eggenstein-Leopoldshafen, Germany}
\author{Dirk Witthaut}
\affiliation{Forschungszentrum J\"ulich, Institute for Energy and Climate Research - Systems Analysis and Technology Evaluation (IEK-STE), 52428 J\"ulich, Germany}
\affiliation{Institute for Theoretical Physics, University of Cologne, 50937 K\"oln, Germany}
\author{Johannes Kruse}
\affiliation{Forschungszentrum J\"ulich, Institute for Energy and Climate Research - Systems Analysis and Technology Evaluation (IEK-STE), 52428 J\"ulich, Germany}
\affiliation{Institute for Theoretical Physics, University of Cologne, 50937 K\"oln, Germany}

\begin{abstract}
The energy transition introduces more volatile energy sources into the power grids. In this context, power transfer between different synchronous areas through High Voltage Direct Current (HVDC) links becomes increasingly important. Such links can balance volatile generation by enabling long-distance transport or by leveraging their fast control behavior. Here, we investigate the interaction of power imbalances - represented through the power grid frequency - and power flows on HVDC links between synchronous areas in Europe. We use explainable machine learning to identify key dependencies and disentangle the interaction of critical features. Our results show that market-based HVDC flows introduce deterministic frequency deviations, which however can be mitigated through strict ramping limits. Moreover, varying HVDC operation modes strongly affect the interaction with the grid. In particular, we show that load-frequency control via HVDC links can both have control-like or disturbance-like impacts on frequency stability. 
\end{abstract}

\maketitle

\section*{Introduction}
The transition to a renewable energy supply challenges the operation and stability of electric power systems \cite{witthaut2021collective}. Wind and solar power generation are determined by the weather and are thus intrinsically fluctuating and uncertain \cite{anvariShortTermFluctuations2016,collinsImpactsInterannualWind2018}. Hence it becomes increasingly difficult to balance the generation and consumption of electric power \cite{elsner2015flexibilitatskonzepte}. Furthermore, the location of renewable power generation is partly determined by natural conditions and not by consumer needs. For instance, wind turbines are preferably installed near the coast, while the centers of the load may be far away inland, leading to higher transmission needs and grid loads \cite{peschImpactsTransformationGerman2014}. To master these challenges, we need new infrastructures, but also new concepts for power system control and operation  \cite{poollaOptimalPlacementVirtual2017,milanoFoundationsChallengesLowInertia2018}, as well as a better understanding of the complex inter-dependencies of different parts of the energy system. This article contributes to this understanding by a data-based analysis of the interplay of power system control and cross-area power flows. 

Load-frequency control is the central method to balance generation and load on short time scales. Although extreme imbalances, e.g. after contingencies, are rare, system balancing becomes a regular challenge due to large fluctuations, unscheduled events, or forecasting errors \cite{schaferNonGaussianPowerGrid2018,krusePredictabilityPowerGrid2020,kruse2021revealing}. Any imbalance manifests in the grid frequency, which decreases from its reference value in case of scarcity of generation and vice versa \cite{anderson2003,machowskiPowerSystemDynamics2008}. In the Central European power grid, the number of large frequency deviations has been increasing in the last decade such that frequency stability has become a matter of increasing importance \cite{entso-eaisblReportDeterministicFrequency2019}. The power balance is restored by activating different layers of reserves, which are generally activated based on deviations of the grid frequency
Primary control or frequency containment reserve (FCR) is activated within seconds and can roughly be described by a proportional control law. Secondary control or frequency restoration reserve (FRR) is activated within minutes and can roughly be described by an integral control law.

Power flows between different synchronous areas play an increasingly important role due to the energy transition. The increasing demand for long-distance power transmission is covered by a variety of grid extension projects \cite{Netzentwicklungsplan}. This includes in particular new high-voltage direct current (HVDC) links within and between different synchronous areas. Within the last decade, several new links have been established in Europe, and further ones are planned or under construction. Links from and to the Scandinavian power grid are particularly promising due to the large resources of (controllable) hydropower and pumped hydro storage in the Nordic area \cite{tellefsen2020norwegian}. 
The operation of such cross-area HVDC links is mainly determined in advance on the market. However, HVDC links and converter stations can also be used in load-frequency control, either optimizing the frequency in both areas or only on one site \cite{entso-HVDCLinksOperationReport}. Then, power flows are adjusted on short terms and thus deviate from the previously scheduled values, which is why unscheduled cross-area flows are inherently connected to frequency control through HVDC links. Until now, pilot projects and simulation studies have demonstrated the additional use of frequency control via HVDC links between synchronous areas \cite{jingyahuangHVDCbasedFastFrequency2017, langwasserEnhancedGridFrequency2020, dijokasFrequencyDynamicsNorthern2021, dehaanStabilisingSystemFrequency2016}. The European network of transmission system operators (ENTSO-E) currently revises the standards for the application of HVDC links within the two implementation projects for coordination of aFRR and mFRR throughout Europe, PICASSO and MARI~\cite{picasso, mari}.

In spite of these developments, cross-area flows on HVDC links can also have unexpected and even unfavorable effects on frequency quality, especially on the Rate of Change of Frequency (RoCoF). Firstly, scheduled power flows on HVDC links can be ramped up or down very fast, which can lead to deterministic frequency deviations \cite{entso-eaisblReportDeterministicFrequency2019}. As fast HVDC ramps cause RoCoFs beyond the systems' control capabilities, technical studies even suggest ramp limits \cite{entso-HVDCLinksOperationReport}. While the effect of fast generation ramps on frequency quality is well-known \cite{weissbachHighFrequencyDeviations2009, kruse2021revealing}, the effects of scheduled HVDC ramps are still under investigation \cite{entso-eaisblReportDeterministicFrequency2019}. Secondly, ramp limits and other technical parameters limit the effect of HVDC links on the grid frequency \cite{jingyahuangHVDCbasedFastFrequency2017}. Here the question arises, how such constraints affect frequency deviations in reality. Thirdly, frequency control on one side of an HVDC link can act as a power disturbance on the other side \cite{langwasserEnhancedGridFrequency2020}, which might cause unfavorable effects on frequency stability. Understanding expected as well as unfavorable effects in historic operational data of HVDC links and grid stability is therefore of great importance. However, multiple factors affect grid frequency deviations in large synchronous areas thus making it hard to isolate the interaction with HVDC power flows by simple data analysis, such as correlation studies \cite{kruse2021revealing}.

In this article, we use eXplainable Artificial intelligence (XAI) for a data-based analysis of HVDC operation, frequency stability, and their interrelations. Machine Learning (ML) allows us to model and disentangle different effects in the data \cite{hastieElementsStatisticalLearning2016}, while methods from XAI enable us to gain insights into the dependencies identified by the model \cite{barredoarrietaExplainableArtificialIntelligence2020}. In particular, we use SHapley Additive exPlanations (SHAPs), which have highly desirable mathematical properties \cite{lundbergUnifiedApproachInterpreting2017,lundbergLocalExplanationsGlobal2020a}. As a case study, we apply these methods to HVDC flows between the major synchronous areas within Europe. Using five years of publicly available data \cite{transnetbwfreqdata, nationalgridfreqdata, fingridfreqdata, ENTSOETransparencyPlatform}, we extract the interactions between HVDC flows and the grid frequency, both for the entire grid as well as for individual HVDC links. 

The article is organized as follows. First, we describe our methodology, which we then apply to two different models.
The stability model aims to predict indicators for frequency stability from various techno-economic features. We focus on the impact of cross-area flows which are extracted via SHAP values. Then, we introduce a second model, the flow model, that targets unscheduled cross-areas flows on individual links, including frequency stability indicators as features. This approach reflects the fact that interactions between flows and frequency can be both-way; in particular, frequency deviations may cause unscheduled flows due to frequency control. Finally, we close our study with a discussion.

\section*{Explainable machine learning for the analysis of complex energy systems}
\label{sec:methods}

We aim to extract the relation between frequency stability and the flows on HVDC lines, that connect 
different synchronous areas within the European power system. Frequency stability depends on a variety of drivers that display complex interactions \cite{kruse2021revealing} such that it is extremely difficult to single out the role of one or even a few key influence factors based on ordinary correlation analysis. Instead, we need a method that can disentangle the contributions of different factors in a mathematically consistent way. Machine learning provides powerful tools to model such complex dependencies. We refer to our code \cite{github} and data \cite{zenodo} for details on our model implementation and data preparation that go beyond the main text.

\begin{figure*}[tb]
    \centering
    \includegraphics[width=0.45\textwidth]{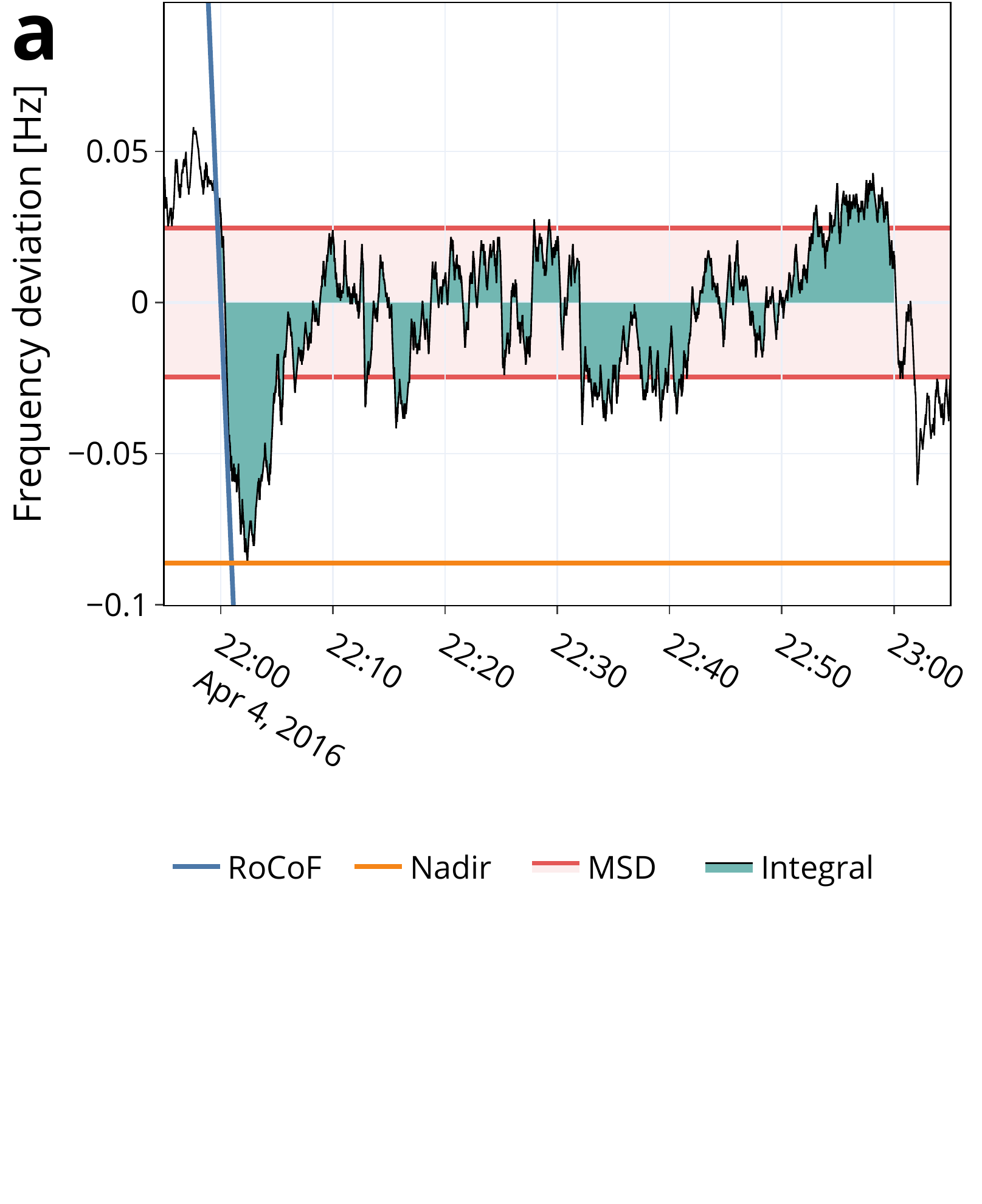}
    \includegraphics[width=0.45\textwidth]{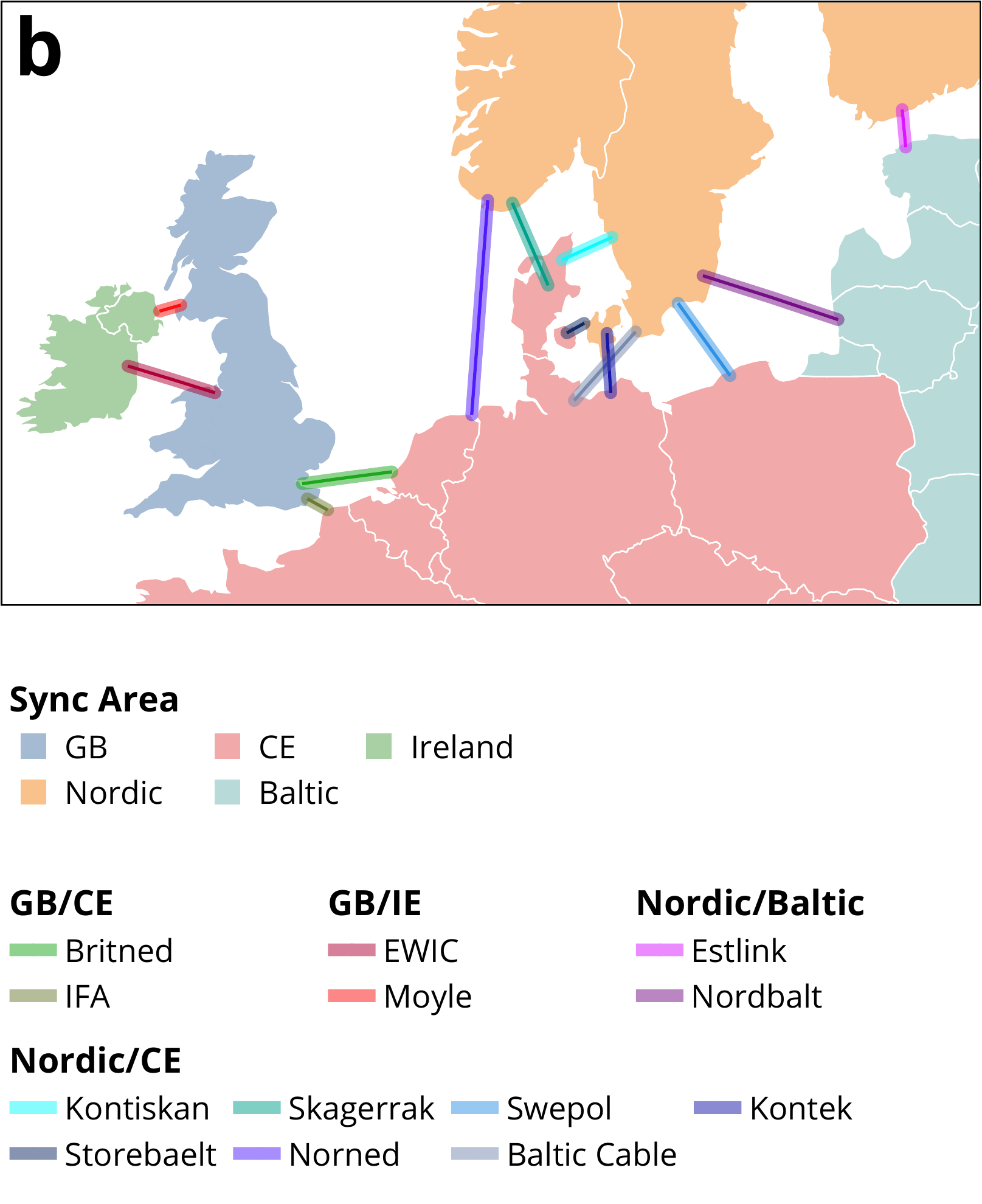}
    \caption{
    (a) Indicators of frequency stability in hourly resolution. The RoCoF is the slope of the frequency at the beginning of the hour while the Nadir is the largest frequency deviation within one hour. We also considered the Mean Square Deviation (MSD) and the integral for each hourly frequency trajectory. (b) Map of northern Europe including the HVDC links we considered in this study. The differently colored regions reflect the five synchronous areas adjoining these HVDC connections.
    }
    \label{fig:indmap}
\end{figure*}

\subsection*{Quantifying frequency stability}
Frequency stability is quantified by four indicators (Fig.~\ref{fig:indmap} (panel a)), which are evaluated on an hourly basis reflecting the basic time interval for electricity trading and scheduling. All indicators are extracted from frequency time series $f(t)$ with a 1-second resolution obtained from  \cite{transnetbwfreqdata, nationalgridfreqdata, fingridfreqdata}. First, the Rate of Change of Frequency (RoCoF) is defined as the gradient of $f(t)$ at the beginning of the hour after the dispatch has been adapted. Second, we consider the integral of $f(t)$ which reflects a sustained frequency deviation and thus a sustained power imbalance. Finally, we consider the largest deviation from the reference frequency during the hour, the nadir, and the mean square deviation. Details are described in ref.~\cite{kruse2021revealing}.

\subsection*{HVDC cross-area flows}
This study investigates the relations of non-embedded HVDC link operation to frequency stability. Therefore, we included the time series of scheduled commercial exchanges and physical flows in the analysis, which was neglected in a previous model~\cite{kruse2021revealing}. The ENTSO-E transparency platform provides the in and outflows between two bidding zones, control areas, or countries ~\cite{ENTSOETransparencyPlatform}.

To include HVDC flows in our models, we engineer two different datasets with different purposes. 

First, we use an aggregated dataset to investigate the impact of flows on the frequency stability indicators. For each synchronous area, we construct a time series summarizing the net inflow from each neighboring synchronous area. 

Second, we use a link-resolved dataset to investigate how the flows depend on frequency deviations. We construct time series from the transmission data between two bidding zones or control areas, which we can attribute to a specific HVDC link or link group. Time periods where links are malfunctioning or unavailable are excluded from the analysis as we want our models to explain the day-to-day behavior. To this end, we utilized the transmission grid unavailability data on the ENTSO-E transparency platform \cite{ENTSOETransparencyPlatform}. We found that omitting all data from listed time periods shorter than two months yielded a good trade-off between eliminating exceptional behavior while maintaining sufficient data for training and testing.

In addition, we constructed time series of unscheduled flows by subtracting physical flows from scheduled commercial exchanges for both datasets. 

Overall, we included transmission data for 13 HVDC links connecting five different synchronous areas, as illustrated in Fig.~\ref{fig:indmap} (panel b).
The transmission data between Belgium and the United Kingdom was generally omitted as the Nemo Link was just commissioned towards the end of the period considered in this paper.

\subsection*{Stability model and flow model}
To investigate the relationship between frequency deviations and HVDC flows, we used two models that differ in their inputs and targets. The stability model, which we will look at first, targets the four frequency indicators introduced above for the synchronous power grids of Great Britain (GB), Northern Europe (Nordic), and Continental Europe (CE) by using multiple techno-economic features. The set of features contains the aggregated inflows from other synchronous areas as described above and several features characterizing generation and load. As in ref.~\cite{kruse2021revealing}, we included time series of day-ahead load forecast, day-ahead wind and solar forecast, day-ahead electricity prices, day-ahead scheduled generation per production type as well as the actual load and generation per type. All time series were engineered from publicly available day-ahead and ex-post data obtained from the ENTSO-E transparency platform \cite{ENTSOETransparencyPlatform}. As we want to model hourly frequency indicators, the time series with higher time resolution were downsampled to hourly resolution by taking hourly averages.  
We aggregated the time series for each synchronous area following the procedure in ref.~\cite{kruse2021revealing}.
Beyond the aggregated features we engineered ramp features (gradients from time $t-1$ to $t$), forecast errors (day ahead - actual), and an inertia proxy inspired by ref.~\cite{ulbigImpactLowRotational2014}.

The second model, the flow model, then targets the unscheduled flows of individual HVDC links. We use one model class for each of the four border regions, which differ in their input feature set. 
For links connecting GB and CE as well as CE and Nordic, data from both terminal sides are available. The feature sets contain all time series on generation, load, and prices from both sides as described for the stability model. In addition, the feature set includes the planned commercial exchange and the four stability indicators from both terminal sides.
For links connecting the Nordic and the Baltic grid as well as the GB and the Irish grid, only data from one terminal side is included due to the unavailability of data from the Irish and the Baltic grid.

\subsection*{Model training, testing, and explanation}
We used the LightGBM framework \cite{ke2017lightgbm} to train gradient boosted trees to our datasets consisting of hourly time series for the years 2015-2020. Gradient Boosted Trees are considered among the highest performing methods for tabular datasets~\cite{chenXGBoostScalableTree2016} while they also allow for fast and efficient computation of SHAP values which enable us to explain the model~\cite{lundbergLocalExplanationsGlobal2020a}. Before training, we shuffled our data and split it into a training set (64\%), a validation set (16\%), and a test set (20\%). To optimize the hyperparameters of each model we conducted a grid search combined with a 5-fold cross-validation on the training set while applying early stopping of the boosting rounds utilizing the validation set. Subsequently, we tested the model performance on an unseen test set. 

Finally, the model is interpreted and analyzed in terms of SHAP values which allow us to disentangle each model prediction into the contribution of the individual features~\cite{lundbergLocalExplanationsGlobal2020a}. Assume that we have a set of $n$ features. Given the feature values $x_1,\ldots,x_n$, the model yields the output $f(x_1,\ldots,x_n)$. Using the SHAP framework, this prediction can be attributed additively to the individual features $j \in \{1,\ldots,n\}$. That is, the output can be written as a sum 
\begin{equation*}
   f(x_1,\ldots,x_n) = \phi_0(f) + \sum_{j=1}^n \phi_j(f, x_1,\ldots,x_n),
\end{equation*}
where the $\phi_j$ is the SHAP value of the $j$th feature and the base value $\phi_0$ is the expected value of $f$. 

However, SHAP values are not limited to the explanation of individual inputs, but can also be used to interpret global feature importances and model behavior by combining many local explanations. The absolute value of a feature's SHAP value reflects how much the model relies on this feature for the corresponding input. Therefore, we can quantify global feature importance within one model by averaging over a feature's SHAP values for all inputs. We then obtain a normalized feature importance by dividing the feature contribution of one feature $k$ by the total sum of all feature contributions: 
\begin{equation*}
    \frac{\left< |\phi_k (f, x_1, \ldots, x_n) |\right>_\text{inputs}}{\sum_{j=1}^n  \left<|\phi_j  (f, x_1, \ldots, x_n) |\right>_\text{inputs}}.
\end{equation*}

Consequently, a feature importance of one would imply that the model solely relies on this single feature while a zero feature importance implicates that the model does not consider the feature at all. Moreover, feature importances for a single model sum to one. We also utilize SHAP dependency plots, which are able to provide much deeper insights than classical dependence plots, especially when SHAP interactions are also taken into account. For instance, SHAP dependence plots are capable of providing much deeper insights than classical dependence plots, especially when SHAP interactions are also taken into account. Nevertheless, it is important to keep in mind that SHAP in its own right is not able to unveil any causal relationship. Therefore any results have to be interpreted using domain knowledge to yield meaningful knowledge discovery.

\section*{Modeling and understanding frequency stability indicators}
\label{sec:freqmodel}
We first consider the stability models, which predict the four different indicators for power systems frequency deviations. We focus on the role of the features describing cross-area flows via HVDC links, which had been discarded in previous models in ref.~\cite{kruse2021revealing}. 

\begin{figure*}
    \centering
    \includegraphics[width=0.95\textwidth]{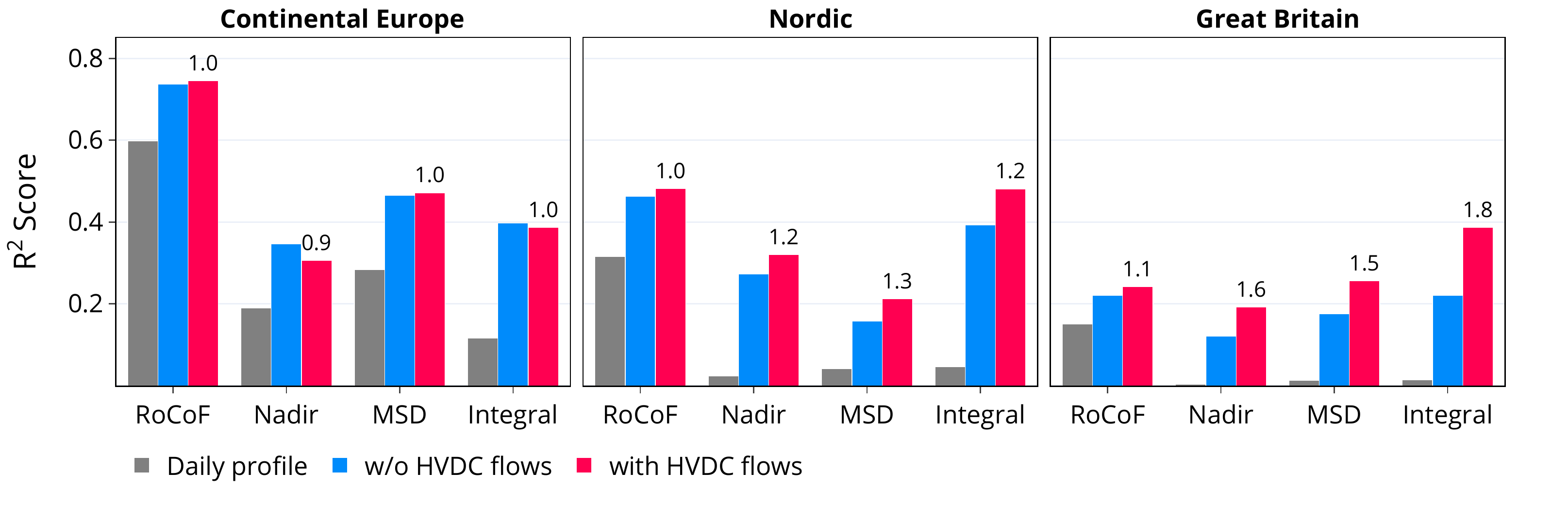}
    \caption{
    Including HVDC transmission data improves the modeling of frequency stability indicators. For three European synchronous areas, we present Machine Learning (ML) models that predict frequency stability indicators from techno-economic features such as generation per type, load, and day-ahead electricity prices. We measure their performance by the $R^2$ score, which quantifies the share of variance explained by the model (colored bars). As a benchmark, we also provide the $R^2$ score for the daily profile predictor, which predicts the targets purely based on their daily average evolution (grey bars). The ML models outperform the benchmark by a large margin, showing the overall importance of techno-economic features for frequency stability. Compared to previous models without HVDC features (cf.~\cite{kruse2021revealing}), the inclusion of cross-area flows improves the performance by a factor of up to 1.8 (indicated by the numbers above the red bars). The benefits of including HVDC flows are particularly large in GB and the Nordic area.
    }
    \label{fig:fmodel-performance}
\end{figure*}

\subsection*{Importance of HVDC flows for frequency stability}
We start our investigation by looking at the overall importance of HVDC flows for frequency stability. The importance is quantified in terms of ML models, comparing the performance with and without the HVDC flows in the feature set. The performance is measured by the R$^2$-score, which reflects the share of variance explained by the model, and evaluated for all four frequency indicators and three different synchronous areas (Fig.~\ref{fig:fmodel-performance}).

We find that the importance of cross-area flows is very different for the three major European grids. In Great Britain, the performance of the ML models increases strongly, up to a factor of 1.8, if cross-area flows are included. In absolute terms, however, the performance remains mediocre with values $R^2 \approx 0.4$. In contrast, the performance is substantially higher for the Central European grid and remains largely unaffected when cross-area flows are included or not. This may be attributed to the fact that the Central European grid is much larger (factor of approximately eight \cite{rydingorjaoOpenDatabaseAnalysis2020}), such that HVDC converter stations take a much smaller relative share in generation or load. 

\begin{figure*}[tb]
    \centering
    \includegraphics[width=0.85\textwidth]{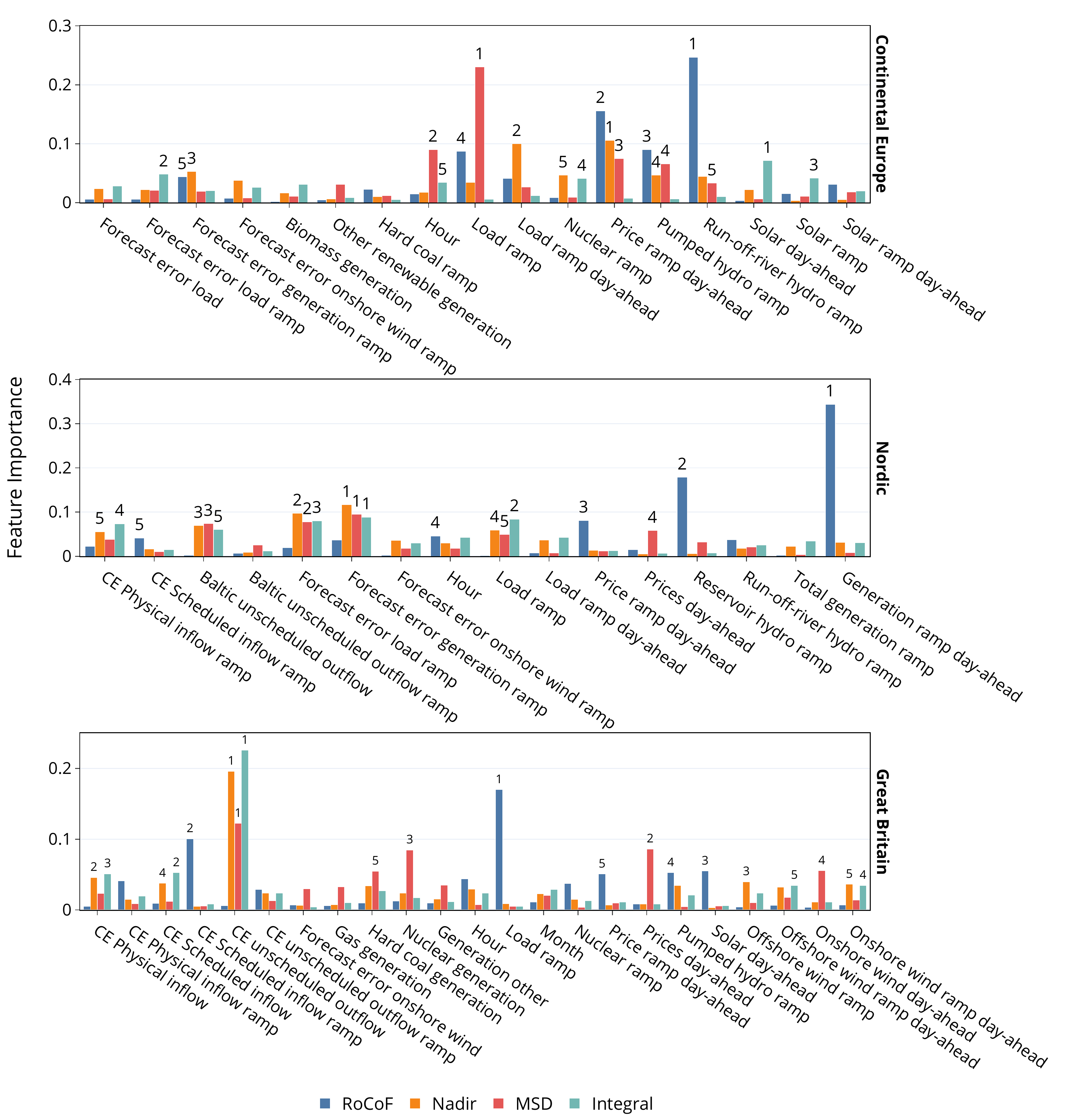}
    \caption{Feature importances of HVDC flows vary among the different synchronous areas. We measure the feature importances in our ML models by the mean absolute SHAP values. 
    The importances for each model are normalized such that they sum to one as described in the main text. For each synchronous area and each indicator, the eight most important features are depicted. In the Nordic and GB areas, the HVDC features are among the most important features, but in CE the model did not identify cross-area flows as highly important. 
    }
    \label{fig:fmodel-imprtance}
\end{figure*}

The general finding is confirmed by a detailed analysis of the feature importances reported in Fig.~\ref{fig:fmodel-imprtance}. In the GB grid, unscheduled, physical, and scheduled flows on HVDC links are among the three most important features for different targets, while HVDC features occur in the Nordic model less prominently. In the CE model, HVDC features are not among the eight most important features. This underlines the importance of HVDC power flows in GB and to a lesser extent the Nordic grid, while HVDC flows have a relatively small impact in CE. 

\subsection*{The effect of scheduled HVDC flows}
Some feature importances reveal relevant effects of scheduled HVDC flows, which are related to deterministic frequency deviations \cite{kruse2021exploring}. Fig.~\ref{fig:fmodel-imprtance} shows a strong impact of scheduled flow ramps on the RoCoF in GB, while the other stability indicators are only weakly affected. This relates to the market-based schedule of flows on HVDC links, which changes mostly in intervals of one hour, thus affecting the hourly RoCoF predominately. Schedule-based ramps on HVDC links introduce temporary power imbalances and thus lead to deterministic deviations in the grid frequency \cite{entso-eaisblReportDeterministicFrequency2019}. This is similar to the deterministic effect of fast, scheduled generation ramps, which are a well-known driver of deterministic frequency deviations \cite{weissbachHighFrequencyDeviations2009, kruse2021revealing}. In the Nordic grid, scheduled flows ramps are ranked as the sixth most important feature for the RoCoF, indicating weak deterministic effects through scheduled HVDC ramps. 

\begin{figure*}[tb]
    \centering
    \includegraphics[width=0.95\textwidth]{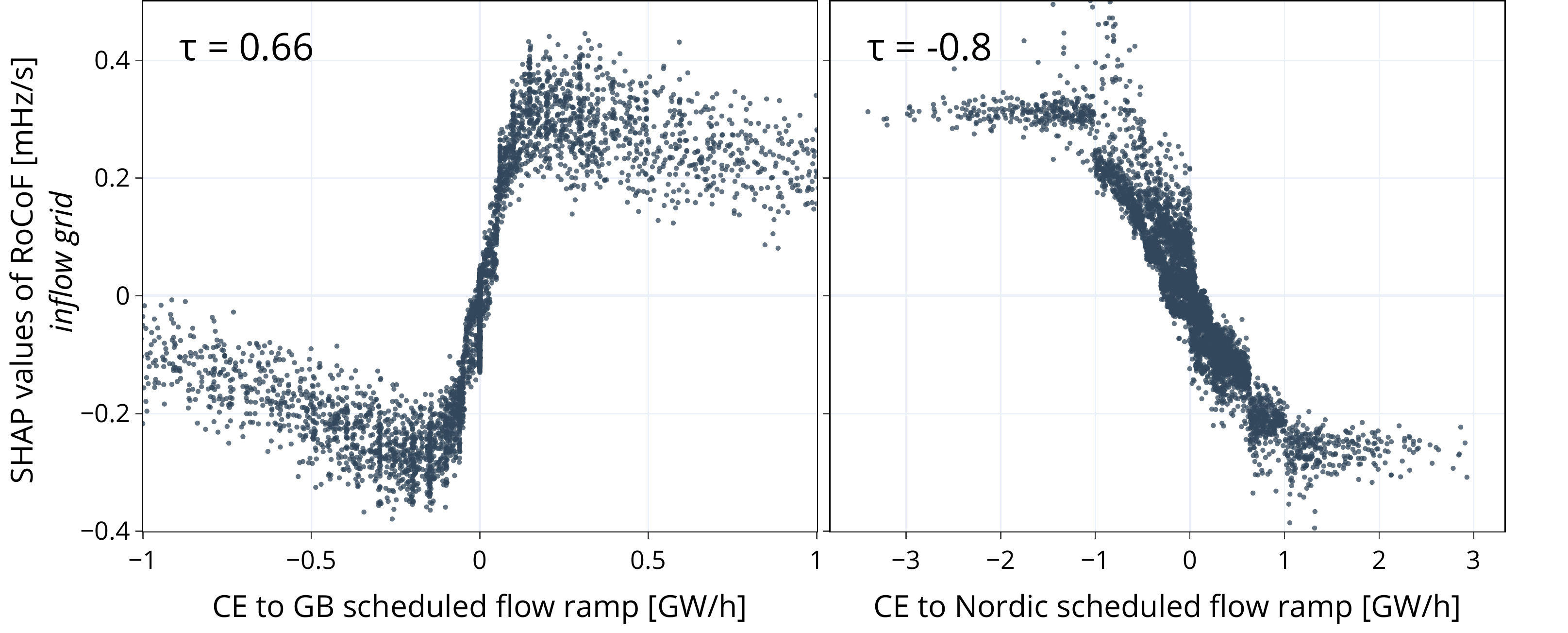}
    \caption{Different effects of scheduled HVDC ramps on the Rate of Change of Frequency (RoCoF). We show the modeled interdependencies between the ramps of scheduled HVDC flow to Continental Europe and the hourly RoCoF in the GB (left) and Nordic (right) synchronous areas. The dependencies have opposite directions, which is confirmed by the Kendall correlation coefficient $\tau$ of the scattered data. We explain these effects with different relative ramp speeds $s$, which we calculate based on the ramp rates $r$ of the HVDC links \cite{entso-eaisblReportDeterministicFrequency2019}. On the CE-GB border, a cable with 1000 MW capacity is allowed to ramp 100 MW/min, i.e., the ramp rate is $r=0.1 / \rm{min}$ (share of full load per minute). On the CE-Nordic border, a cable such as Kontek has 600 MW capacity and Nordic ramp restrictions allow 600 MW per hour, i.e., a ramp rate of $r=0.017 / \rm{min}$. Using the mean $\Delta P$ of the absolute hourly flow changes on the links, we estimate the typical Rate of Change of Power as RoCoP$=\Delta P \cdot r$, which results in different relative ramp speeds $s$ (see main text). These different ramping speeds most probably explain the different effects on the hourly RoCoF.
    }
    \label{fig:rocof}
\end{figure*}

Interestingly, the direction of these ramping effects varies among the grids (Fig.~\ref{fig:rocof}). While increasing scheduled ramps leads to larger RoCoFs in GB, the relation is the opposite in the Nordic grid. Following the discussion in ref.~\cite{kruse2021revealing}, we thus identify the ramping effect in GB as RoCoF-driving, while in the Nordic grid HVDC ramps are RoCoF-offsetting, i.e. larger scheduled ramps are rather related to a RoCoF reduction.

We explain this observation with ramp speeds of HVDC converters relative to the speed of other generation types in the respective synchronous area. We start from typical values of the rate of change of power (RoCoP) for every type. Then we compute the ratio with respect to the fastest RoCoP in the grid, yielding the relative ramp speed $s \in [0,1]$ similar to the procedure in ref.~\cite{kruse2021revealing}. 

In GB, scheduled HVDC ramps are equally fast ($s=0.14$) as other RoCoF-driving generation types such as pumped hydro generation ($s=0.15$), thus causing a step-like behavior, which drives the deterministic frequency deviations. In the Nordic grid, fast reservoir hydro ramps are the dominating RoCoF-driving technology ($s=1$) and inflow ramps from CE are relatively slow ($s=0.04$). This is most probably due to the strict ramping limits on HVDC links imposed by Nordic TSOs \cite{entso-eaisblReportDeterministicFrequency2019}. In this way, HVDC ramps rather smooth the generation curve and dampen the temporary power imbalance, which explains the different dependencies in Fig.~\ref{fig:rocof}. In CE, scheduled HVDC ramping speeds are also very slow ($s=0.03-0.09$) as large generation ramps dominate the overall RoCoP due to the size of the area. Thus, scheduled HVDC ramps are not strongly important for the RoCoF in CE and do not appear in Fig.~\ref{fig:fmodel-imprtance}. 

\begin{figure*}[tb]
    \centering
    \includegraphics[width=0.9\textwidth]{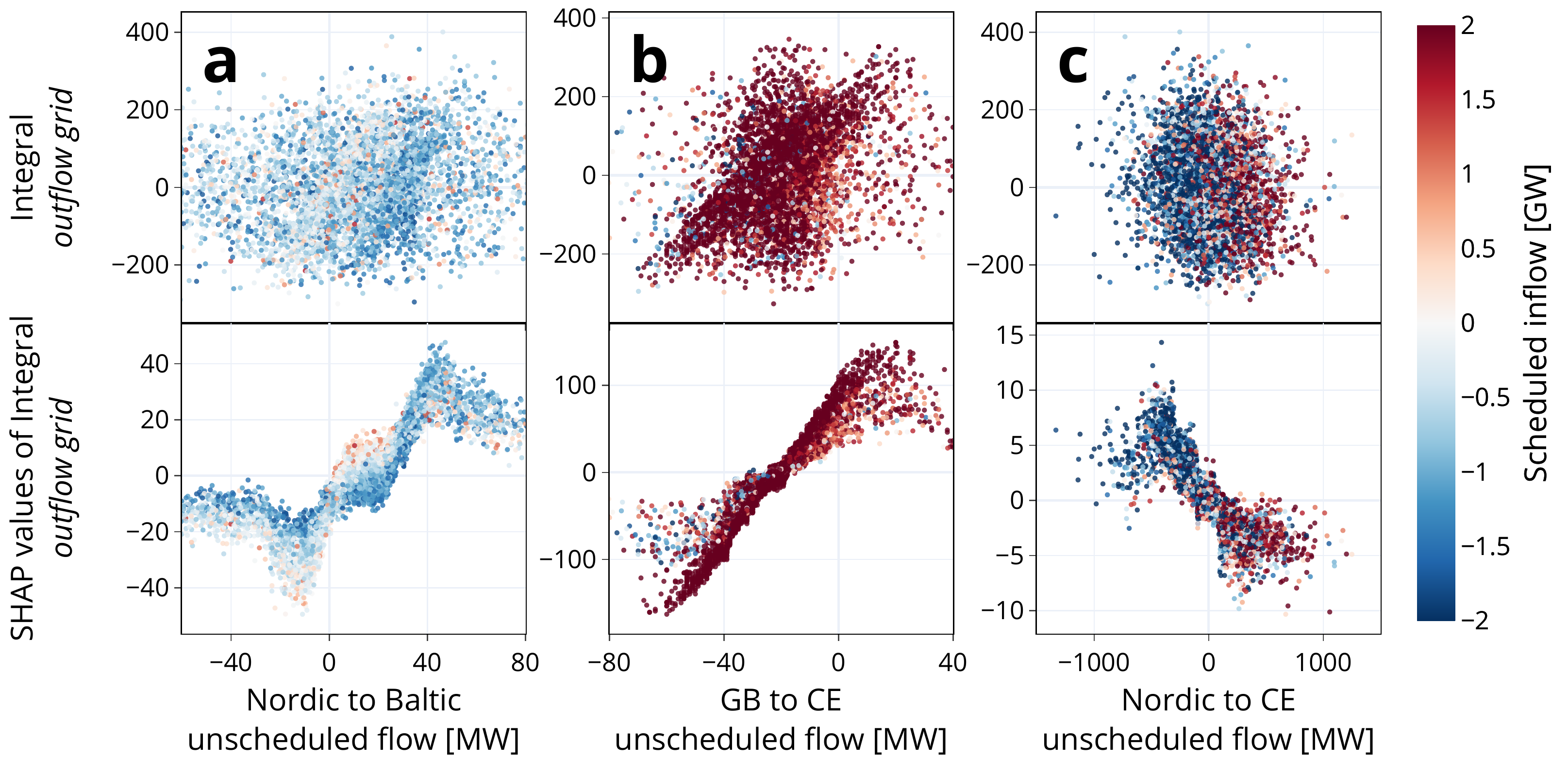}
    \caption{
    Unscheduled flows and the frequency integral show different dependencies among the European grids.
    The model-independent scatter plots (top row) of unscheduled outflows and frequency integrals do not exhibit clear dependencies. Therefore we display SHAP dependency plots (bottom row), which depict the relationship extracted by the stability model thus isolating the effect of a single feature. 
    The integral in the Nordic (a) and the GB area (b) show a positive relation to unscheduled flows, which corresponds to  
    the effect of frequency control, while the CE unscheduled flows have a negative effect on the Nordic integral (c) thus showing a disturbance-like behavior. 
    The scheduled flows (color code) demonstrate that Nordic-CE and GB-CE flows are mostly uni-directional, while CE-Nordic flows exhibit both directions (see main text for discussion).
    }
    \label{fig:fmodel-dependency}
\end{figure*}

\subsection*{The effect of unscheduled HVDC flows}
In addition to scheduled flows, unscheduled HVDC flows, i.e., deviations from the market-based schedule, play an important role in frequency deviations. Figure~\ref{fig:fmodel-imprtance} reports CE unscheduled flows to be the most important feature for three stability indicators in GB. In the Nordic grid, unscheduled flows and the related physical flows are also of high importance. 
The importance of unscheduled flows in frequency dynamics might relate to the effect of unforeseen outages on the HVDC links, but these are rare events \cite{entso-NordicBalticHVDCStatistics}, which cannot strongly influence the dependencies within our model. Another, more realistic explanation for the importance of unscheduled flows is load-frequency control, which can be applied through HVDC links \cite{entso-HVDCLinksOperationReport} and always introduces an unscheduled change in the power flow on the link.

To further investigate this hypothesis, we exemplarily examine the effect of unscheduled flows on the frequency integral for three different HVDC connections 
(Fig.~\ref{fig:fmodel-dependency}). We focus on the frequency integral, as it most closely reflects the need for frequency control via HVDC links in our dataset. The data only includes hourly averaged power flows, so that only net hourly balancing actions through HVDC links are recorded. These are most closely related to the frequency integral as it reflects the net power imbalance within the hour.
Remarkably, the scatter plots of flows and integral values (top row Fig.~\ref{fig:fmodel-dependency}) do not identify clear dependencies. In contrast, the SHAP dependency plots (bottom row) isolate the effect of the individual feature and show distinct relations, thus confirming the advantages of XAI methods over simple correlations analysis (cf.~\cite{kruse2021revealing}).

The observed effects of unscheduled HVDC flows vary among the grids, showing either a control- or a disturbance-like behavior. A control-like effect resembles the behavior of a frequency controller, i.e., positive frequency deviations lead to positive outflows to balance the oversupply within the system. Contrary, a disturbance-like effect entails a negative dependency, where a positive frequency deviation is triggered by a sudden negative outflow, i.e., the sudden inflow of power into the system. In Fig.~\ref{fig:fmodel-dependency} (bottom row), unscheduled outflows to the Baltic area have a positive relation to the Nordic frequency integral (panel a), which is similar to CE unscheduled flows in GB (panel b). Thus, these effects on the Nordic and British frequency show a control-like behavior. In contrast, unscheduled outflows from the Nordic grid to CE (panel c) show a negative dependency, which resembles a disturbance-like effect. Interestingly, also the physical flows (color code) differ strongly among the three examples in Fig.~\ref{fig:fmodel-dependency}, which we will discuss within the next section.

Finally, we note that unscheduled HVDC flows are the result of explicit operation strategies by the adjoining TSOs,  which are in control of the DC power flow on the links \cite{entso-HVDCLinksOperationReport}. To further understand the varying effects of unscheduled HVDC flows, we thus have to examine the individual operation modes on different links between the synchronous areas. 

\section*{Modeling and understanding unscheduled flows between synchronous areas}
\label{sec:flowmodel}
Our data-based analysis shows a strong relation between unscheduled flows and frequency stability. To further understand this interaction, we now switch the perspective and investigate unscheduled flows on individual HVDC links as targets. For this purpose, we use our second model, the flow model, which predicts unscheduled flows on single HVDC links based on techno-economic features and the frequency indicators. 

\begin{figure*}[tb]
    \centering
    \includegraphics[width=0.95\textwidth]{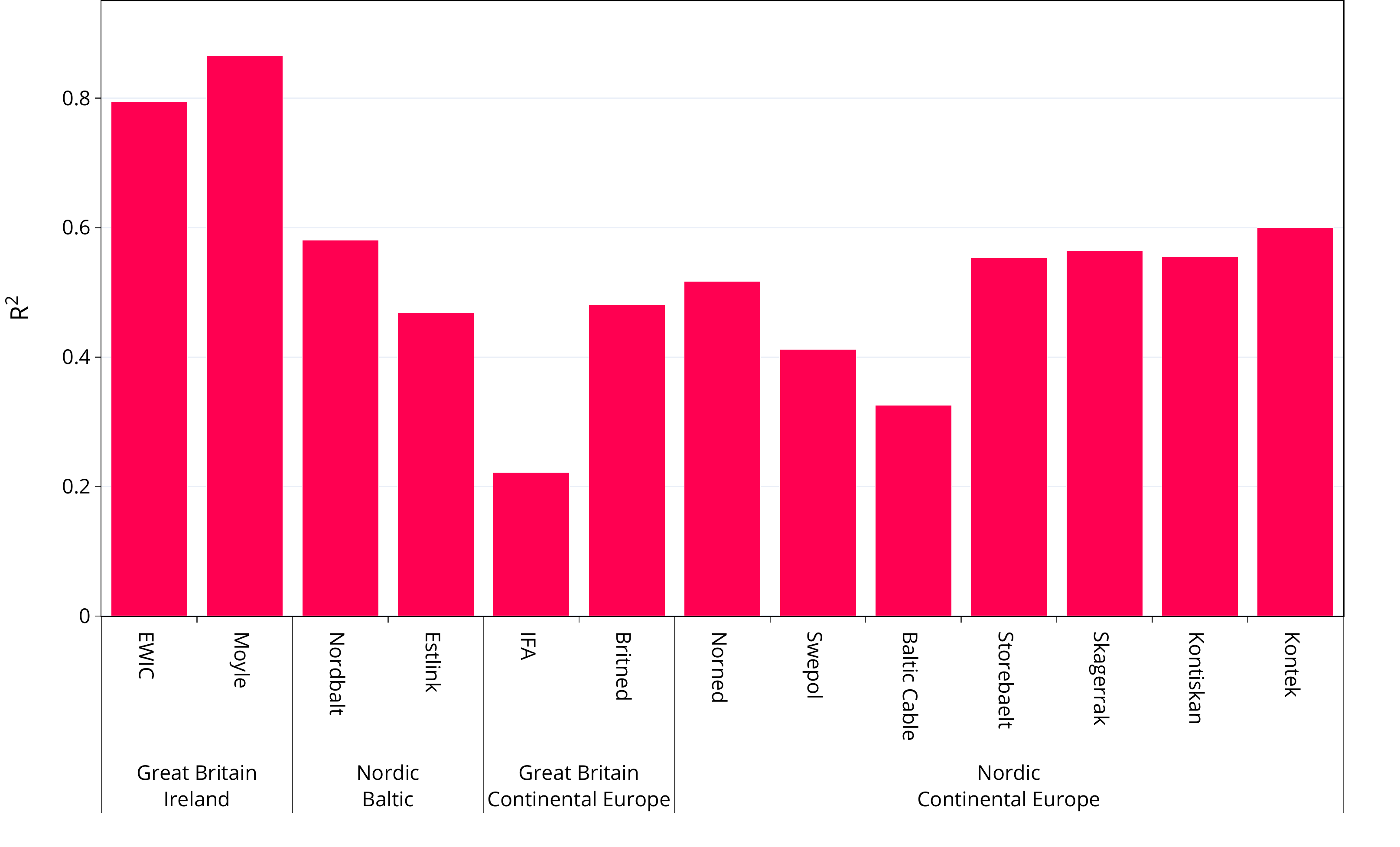}
    \caption{
    With our flow model, we predict the unscheduled flows on single HVDC links between different synchronous areas using techno-economic features such as prices, loads, or generation per type as well as frequency stability indicators.
    The models successfully predict unscheduled flows with performances similar to those of the stability model (cf.~Fig.~\ref{fig:fmodel-performance}).}
    \label{fig:flowmodel-performance}
\end{figure*}

\subsection*{Importance of power imbalances for unscheduled flows}
The flow model can explain a large share of unscheduled flows on various HVDC links (Fig.~\ref{fig:flowmodel-performance}). As a recap, the flow model uses features from the stability model, such as the load or scheduled HVDC flows, as well as the frequency indicators from areas that are available in our data set. Based on these features, we achieve similar performances as in the stability models ($R^2 \sim$ 0.25...0.85), which are acceptable values given the stochastic nature and our limited access to all influencing variables.  Among the feature set, the role of the frequency integral is particularly interesting, as it closely relates to the control need within the hour.  

The frequency integral plays a major role for some HVDC links, which can be related to load-frequency control delivered by these links.
We first focus on the feature importance, which is indicated in Fig.~\ref{fig:integral_importance} on the y-axis, and the feature rank, represented by black numbers. 

The Britned link is most prominent, as the British frequency integral is the most important feature in the respective model (blue color). The Nordic integral (orange color) is ranked among the ten most important features for Estlink, Kontiskan, and Storebaelt.  In contrast, the Nordned link and particularly the Baltic cable show very low feature importances of the integral.

Our findings are fully consistent with operation modes of HVDC links reported by the transmission system operators and the ENTSO-E. Kontiskan, Storebaelt, and Estlink \cite{NordicSOA2019-LFCannex, NordicSOA2006} as well as Britned \cite{entso-HVDCLinksOperationReport} are used in load-frequency control.
As the time series for physical and unscheduled flows correspond to the hourly average, the integral is the relevant feature of the frequency time series.
The situation is very different for the Nordned link, which is not used in load-frequency control as reported in ref.~\cite{entso-HVDCLinksOperationReport}. Consistently, the frequency integral exhibits very low feature importance in the model.

Our studies complement and augment the sparsely available public information on HVDC operation. First, public information is entirely lacking for some of the HVDC links. For instance, we are not aware of any official report describing the operation of the Baltic cable. Our results show that this connection is likely not used for load-frequency control.
Second, most available documents report only the general participation in load-frequency control without further details. Our results allow us to quantify this dependency as we will discuss in the following section.

\subsection*{Control-like and disturbance-like effects}
\begin{figure*}[tb]
    \centering
    \includegraphics[width=0.9\textwidth]{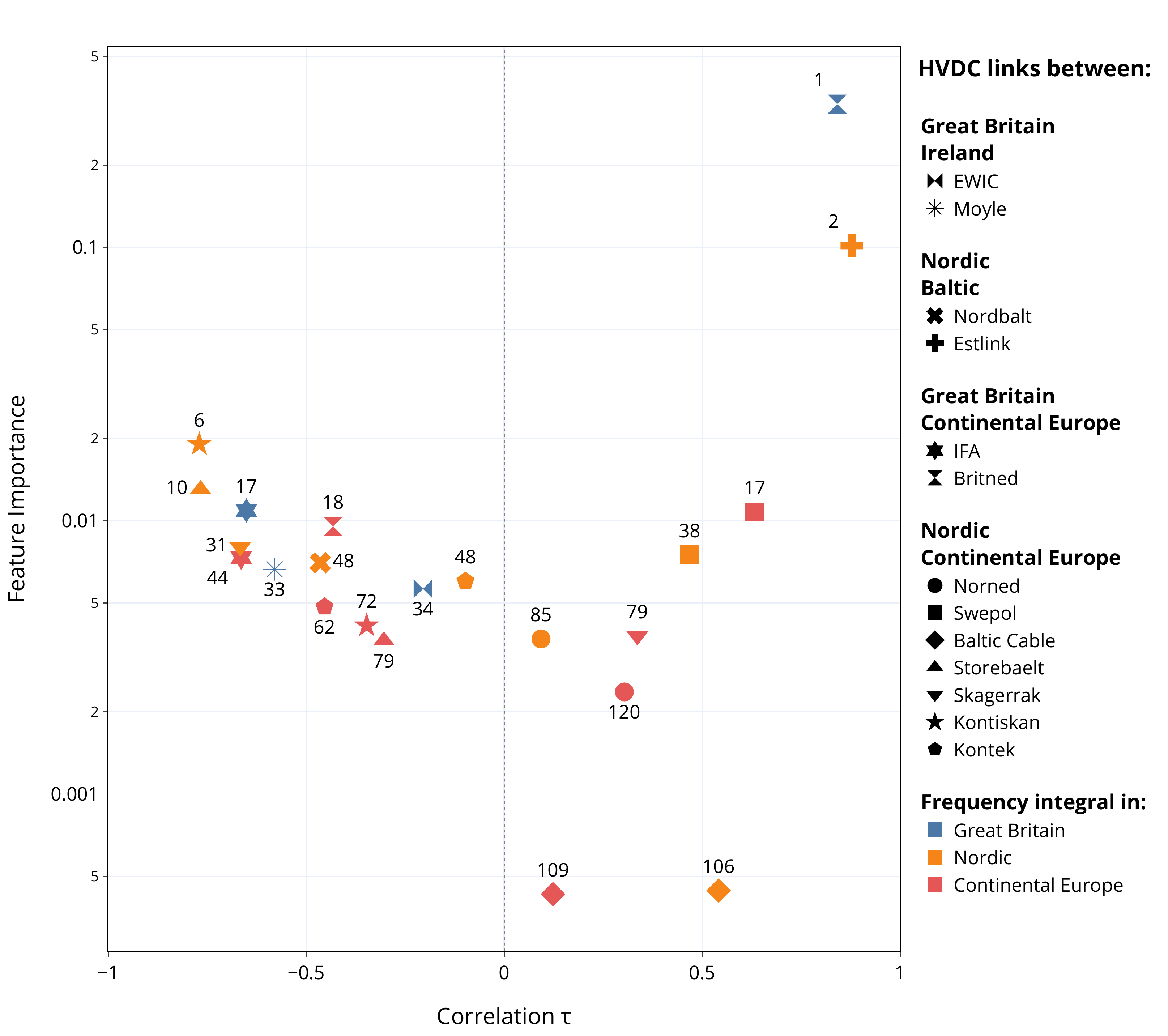}
    \caption{Relation between frequency integrals and unscheduled flows indicates different HVDC operation modes. We show the feature importance of the frequency integrals, which is the mean absolute SHAP value normalized by the sum of all such values, i.e. it becomes one if the model uses only this single feature. The numbers indicate the feature rank within the respective model. For a specific area, the values $\tau$ depict the correlation between the SHAP values of a frequency integral and the unscheduled outflows from this area, i.e., disturbance-like relations appear on the left side and control-like interactions on the right side. The strong control-like effects in the Britned and Estlink model probably relate to their uni-directional control scheme, while the weaker disturbance-like effects for Kontiskan and Storebaelt can be a result of the bi-directional application of frequency control. 
    }
    \label{fig:integral_importance}
\end{figure*}

Turning from feature importances to dependencies, we reveal control-like and disturbance-like interactions between unscheduled flows and frequency stability (Fig.~\ref{fig:integral_importance}). The correlation $\tau$ between the SHAP values of the frequency integral and the unscheduled flows quantifies the direction of this dependency (similar to the correlation in Fig.~\ref{fig:rocof}). 
We note that we define the direction of the unscheduled flows such that positive values correspond to an outflow of electric energy. This holds for all connections such that a flow is counted differently for the two terminal sides. For instance, an unscheduled flow from CE to GB would be counted as positive for CE and as negative for GB.

In that way, control-like dependencies appear in the right half of the plot ($\tau>0$), where positive frequency deviations lead to more outflows, while disturbance-like effects lie in the left half ($\tau<0$). We now discuss these effects for the four links exhibiting the highest feature importances of the integral, i.e., Britned, Estlink, Konstiskan, and Storebaelt.

Control-like effects are most pronounced for Britned and Estlink, which show high feature importance of the GB and the Nordic integral and a strong positive correlation. In contrast, the CE integral in the Britned model is negatively correlated thus showing a disturbance-like effect. (Note that the Baltic integral is not included in the Estlink model due to missing data). These observations are consistent with reports and TSO agreements. Britned is used only for uni-directional control, i.e. only for frequency support in GB \cite{entso-HVDCLinksOperationReport}, and Estlink is also dominantly used by the Nordic side for load-frequency control, according to the 2019 System Operator Agreement (SOA) \cite{NordicSOA2019-LFCannex}. 
Thus, on the GB and Nordic sides, we only have a control effect, which introduces a positive dependency between frequency deviations and unscheduled outflows. On the other side of the link, the control actions behave like a disturbance, which explains the negative correlation of the CE integral in the Britned model.

Disturbance-like effects are most pronounced for Kontiskan and Storebaelt, which exhibit a negative correlation for the Nordic integral, but with relatively low feature importance.  In contrast to Britned, the integral on \textit{both} sides of the Kontiskan and Storebaelt links show a negative dependency (red and orange symbols), although the CE integral exhibits even lower feature importance. Reports and TSO agreements suggest, that both links are used bi-directionally for control, i.e., both sides can receive and supply frequency support \cite{NordicSOA2019-LFCannex, NordicSOA2006}. This might explain both the low feature importance as well as the observed dependencies. Both areas experience the control and the disturbance-like effect, which might cancel each other out. This might partly explain the lower importance of the frequency integral for both links. However, the disturbance-like effects dominate thus yielding the observed negative correlation for both the Nordic and the CE integral. One explanation can be the control delay, particularly if links are used for slow tertiary control (mFRR), which is slower than primary and secondary control. On the receiving side, a frequency deviation might trigger a delayed control action that partly falls into the next hour. On the supporting side, the disturbance-like effect of this action appears immediately and thus might introduce a stronger dependency between the hourly features and targets. However, also other effects of HVDC operation might play a role here. 

\begin{figure*}[tb]
    \centering
    \includegraphics[width=0.85\textwidth]{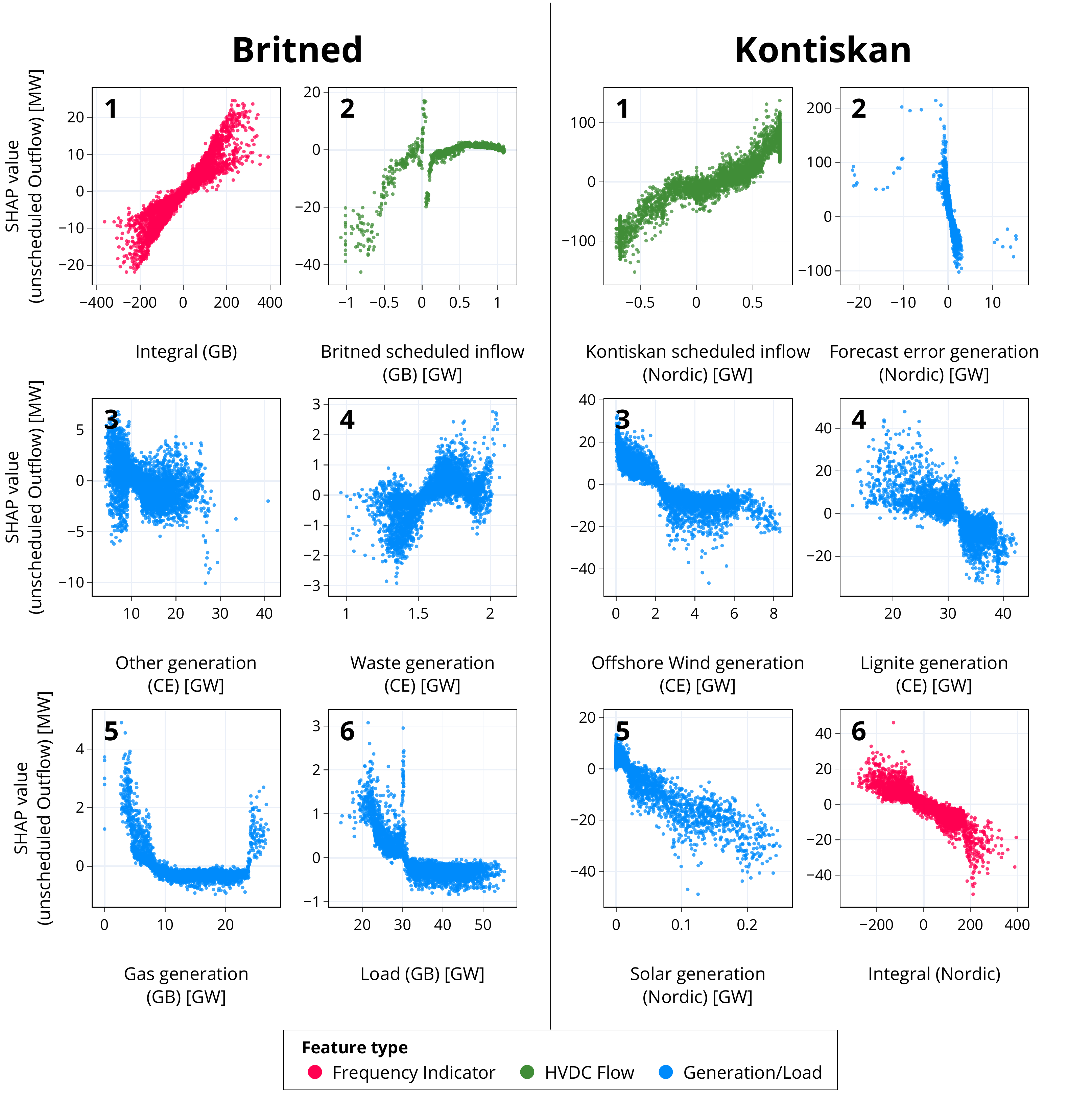}
    \caption{Multiple factors affect the regulation of unscheduled flows on single HVDC links. We show the dependency plots for the first six most important features (measured by mean absolute SHAP values) in the flow models of Britned and Kontiskan as examples. The reference area for unscheduled outflows is GB (for Britned) and the Nordic area (for Kontiskan). While the frequency integral dominates the Britned model, other features such as the scheduled HVDC flows and actual load and generation also affect the unscheduled power flows on the two links. 
    }
    \label{fig:shap_scatter}
\end{figure*}

\subsection*{Other effects of HVDC operation modes}
The frequency integrals, i.e., systematic power imbalances, do not alone describe the unscheduled HVDC flows as demonstrated by their low feature importance for multiple links (Fig.~\ref{fig:integral_importance}). However, most flow models still exhibit a high performance (Fig.~\ref{fig:flowmodel-performance}) thus pointing to other highly predictive features. These effects can relate to other properties of HVDC operation, which we discuss using the Britned and Kontiskan links as examples (Fig.~\ref{fig:shap_scatter}).

The scheduled flows play a major role for both Kontiskan and Britned (green color). For Kontiskan, this feature is even the most important one, with the integral (red color) only following at rank six. Scheduled inflow on Kontiskan to the Nordic area is associated with an increase in the amount of unscheduled outflow. For Britned the impact of scheduled flows is nearly zero if the flows direct to GB, which is the dominating case for this border (cf. the physical flows in Fig.~\ref{fig:fmodel-dependency}). A strong negative impact on the unscheduled outflows is only observed for a few data points with scheduled outflows from GB, i.e., scheduled outflows from GB systematically overestimate the physical outflow. The different effects of scheduled flows on Britned and Kontiskan might relate to different operation modes. On Britned the link overload capacity is used for frequency support \cite{entso-HVDCLinksOperationReport}, such that frequency control can operate nearly independently of scheduled flows, which is consistent with a nearly vanishing impact of scheduled flows for the majority of time steps (Fig.~\ref{fig:shap_scatter}). Only in the rare situation of strong outflows to CE, this situation changes. In contrast, no capacity is reserved for frequency control on Kontiskan \cite{NordicSOA2006}. Control can only be provided if the market-based schedule leaves free capacity in the right direction, such that scheduled inflows to the Nordic area allow for increased unscheduled outflows as depicted by the dependency plot (Fig.~\ref{fig:shap_scatter}). 

If we look back at the stability models and Fig.~\ref{fig:fmodel-dependency}, we can even see indications for these effects using the aggregated flow data.
For the connections Nordic-Baltic and CE-GB, the scheduled flow is oriented in the same direction for most hours of the year as indicated in the color code in Fig.~\ref{fig:fmodel-dependency}. GB mostly imports power via HVDC links,  such that positive control power requires a further increase of the flows beyond the commercially scheduled values. Obviously, this is possible only if capacity is available, which is again consistent with ref.~\cite{entso-HVDCLinksOperationReport} and our interpretation of Fig.~\ref{fig:shap_scatter}. 
In contrast, links between the Nordic and CE areas are used for power flows in both directions.

Other important features are the actual generation of different types, actual load, and forecast errors (blue color). Most of these dependencies exhibit a strong vertical dispersion, which indicates strong interactions with other features \cite{lundbergLocalExplanationsGlobal2020a}. Actual load, generation, and forecast errors directly relate to actual power imbalances, which can explain their importance for unscheduled HVDC flows. These features can indirectly also reflect market situation, e.g., intra-day prices or reserve energy prices, which are not included in the model. These market-based features can also influence unscheduled HVDC flows, as control via these links might only be applied if it is cheaper than other domestic operational reserves.  

All in all, Britned is used for uni-directional control in GB, which in most cases does not depend much on the market-based schedule of the link, thus introducing a strong control-like interaction with the British grid frequency. This strong control effect on Britned most probably explains the control-like dependency, which we observed for the aggregated stability model in Fig.~\ref{fig:fmodel-dependency} (panel b). In contrast, Kontiskan exhibits disturbance-like interactions with the grid frequency on both sides, which probably relates to a bi-directional control setting. Control actions are further constrained by scheduled flows, thus diminishing the importance of the grid frequency for unscheduled flows. Similar observations can be made for other HVDC links between the Nordic and the CE area (Fig.~\ref{fig:integral_importance}), thus explaining the disturbance-like effect observed in our aggregated stability model in Fig.~\ref{fig:fmodel-dependency} (panel c).

\section*{Conclusion and Outlook}

In summary, we revealed important dependencies between non-embedded HVDC operation and grid frequency stability using explainable machine learning. Using our publicly available data set and model \cite{github, zenodo}, we extracted such dependencies both for the entire synchronous areas as well as for individual HVDC links between synchronous areas.

First, we highlighted the important role of HVDC flows to model frequency stability indicators, particularly for the GB and Nordic synchronous areas. Between CE and GB, flow ramps drive the British RoCoF as they are relatively fast in the power grid compared to other generation types. In the Nordic grid, scheduled flow ramps rather offset the hourly RoCoF, which is most probably a result of strict ramp limits. This suggests that the ramp limits imposed by the Nordic TSOs are successful in preventing stability problems through fast HVDC ramps.

Second, we successfully modeled unscheduled flows on individual HVDC links based on techno-economic features, including frequency stability indicators. For two links in the GB and Nordic areas, we highlighted how frequency stability needs drive unscheduled HVDC flows, which is most probably related to their important role in load-frequency control.Meanwhile, unscheduled flows on most other lines seem to be driven by factors not directly related to frequency stability. Importantly, if balancing power is provided from one area to the other through HVDC links, power is lost on the supporting side and the frequency drops. This disturbance-like effect was particularly strong in the Nordic grid, which supplies control power to Continental Europe. In the British grid, the control-like effect dominated, i.e. the British side was stabilized, which probably relates to the fact that balancing is done only on the British side of the link. 

Currently, these varying effects are a direct consequence of the differences in operation concepts of the different HVDC lines, which follow bilateral contracts between the respective TSOs \cite{NordicSOA2019-LFCannex, NordicSOA2006, entso-HVDCLinksOperationReport}. ENTSO-E is currently developing PICASSO and MARI to provide a unified concept for aFRR and mFRR throughout Europe, including the application of HVDC in load-frequency control \cite{picasso, mari}. In this context, effects such as the decrease of frequency quality due to frequency support via HVDC are important for the evaluation of HVDC-based load-frequency control concepts. In the future, data-driven methods, such as our approach, can provide valuable tools for analyzing and monitoring the changes due to new control concepts and regulations. 

\section*{Acknowledgements}
The authors thank Thomas Dalgas Fechtenburg for fruitful discussions.

\section*{Funding} 
J.K. and D.W. gratefully acknowledge support from the Helmholtz Association via the Helmholtz School for Data Science in Life, Earth and Energy (HDS-LEE). B.S. gratefully acknowledges funding from the Helmholtz Association under grant no. VH-NG-1727.

\section*{Availability of data and materials} % DO NOT REMOVE THIS SECTION
Our data set, including the techno-economic features, stability indicators, and HVDC flows as well as the results of our hyper-parameter optimization, is available on Zenodo \cite{zenodo}. The code is accessible on GitHub \cite{github}. 

\section*{Author's contributions}
S.P.: Investigation, Formal analysis, Visualization. B.S.: Supervision, Writing, Conceptualization. D.W.: Project administration, Funding acquisition, Supervision, Writing, Conceptualization. J.K.: Project administration, Supervision, Writing, Conceptualization, Methodology.

\section*{Competing interests}
The authors declare that they have no competing interests.

\bibliographystyle{naturemag}
\bibliography{references}

\end{document}